\documentstyle[sprocl]{article}

\def\Journal#1#2#3#4{{#1} {\bf #2}, #3 (#4)}

\def\NPB{{\em Nucl. Phys.} B}
\def\PLB{{\em Phys. Lett.}  B}
\def\PRL{\em Phys. Rev. Lett.}

\def\be{\begin{equation}}
\def\ee{\end{equation}}
\def\bea{\begin{eqnarray}}
\def\eea{\end{eqnarray}}

\begin{document}

\title{Topological symmetry breaking and the 
confinement of anyons}

\author{P.W. Irwin }
\address{Groupe de Physique des Particules, D\'epartement de Physique,\\
Universit\'{e} de Montr\'{e}al, \\
C.P. 6128 succ. centre-ville, Montr\'{e}al, Qu\'{e}bec, 
H3C 3J7, Canada}

\author{M.B. Paranjape\footnote{permanent address: Groupe de Physique des 
Particules, D\'epartement de Physique,
Universit\'e de Montr\'eal, C.P. 6128 succ. 
centre-ville, Montr\'eal, Qu\'ebec, H3C 3J7, Canada}
}
\address{Departamento de F\'{\i}sica T\'eorica, Facultad de Ciencas, Universidad de Zaragoza\\
Zaragoza, 50009, Spain\\E-mail: paranj@lps.umontreal.ca} 

\maketitle\abstracts{ We study topological symmetry breaking via the behaviour of Wilson and 't Hooft loop operators for the 2+1 dimenesional Abelian-Higgs model with Chern-Simons term.  The topological linking of instantons, which are closed vortex loop configurations, give rise to a long-range, logarithmic, confining potential between electric charges and magnetic flux tubes even though all perturbative forces are short range.  Gauss' law forces the concomitance of charge and magnetic flux, hence the confinement is actually of anyons. }

The Abelian-Higgs model in $2+1$ dimensions with Chern-Simons \cite{jdt} term is a field theoretic model describing anyons\cite{wn}.  In the phase where the $U(1)$ symmetry is good, the charged states are forced by the Gauss law to append an amount of magnetic flux that depends on the coefficient of the Chern-Simons term. Meanwhile in the symmetry broken phase, there exist vortex states, which also due to the Gauss law must append an amount of charge that depends on the coefficient of the Chern-Simons term.  Charge-flux composites, sometimes called cyons\cite{mg}, are well known to satisfy fractional statistics and carry fractional angular momentum\cite{p}, and hence are anyons.

In the Abelian Chern-Simons theory, gauge invariance does not require quantization of the coefficient of the Chern-Simons term.  There are no topologically non-trivial gauge transformations that shift the value of the Chern-Simons integral independently of the background gauge field, as in the non-abelian case\cite{jdt}.  However, on certain compact space-time manifolds with non-trivial background magnetic fields present, even the Abelian Chern-Simons term can be gauge non-invariant.  Initially this was seen as a reason for imposing the quantization of the coefficient, however it was soon realized, that such a non-invariance would simply force a projection of the theory to sectors with trivial magnetic field backgrounds, where the Chern-Simons term is gauge invariant\cite{po}.  

The specific situation considered in Ref. 5 corresponds to the manifold $S^2\times S^1$.  The gauge field is taken as a connection on the one monopole bundle with one flux quantum piercing the $S^2$.  Dynamics  typically dictate that the flux is concentrated in a localized flux tube.   Then there exist non-trivial gauge transformations that wrap around the $S^1$ with an integer winding number for which the Chern-Simons term is not invariant.  The shift of the Chern-Simons term is by an integer $N$, multiplying a factor which is proportional to the coefficient of the Chern-Simons term. The integer $N$, depends on the winding number of the gauge transformation and on the number of flux quanta in the magnetic field. The option that the coefficient of the Chern-Simons term must be quantized so that the overall shift is equal to $2 N \pi $ is too strong and need not necessarily be the case.  Instead the conclusion can be drawn that the one monopole, or indeed, the multi-monopole sector does not exist due to this lack of gauge invariance.  However the zero monopole sector does exist since the Chern-Simons term is perfectly gauge invariant. We find this conclusion reasonable.  

However the point of the present investigation is to examine whether this projection to the sub-sector of zero total flux is no more than an unimportant, unphysical, uninteresting, irrelevant, boundary condition-like constraint.  The simple analogy can be made with the constraint in classical electrodynamics that the total charge on any compact manifold must be zero due to the Gauss law. This constraint, however, has no dynamical effect on local charge, which essentially behaves as if it were unconstrained.  It is a well known in concept in field theory, that the vacuum sector contains all the other sectors.  Indeed, it is possible to imagine that one could construct a flux anti-flux pair, and move the anti-flux very far away, so as to locally construct the one flux sector.  ``Very far away" must be on the one hand large compared to any dynamical scale, but smaller than the radius of the $S^2$, which is always possible for a large enough $S^2$.  Many kinds of impediments can arise which can  impeach the existence of local single flux states.  There can be lack of cluster decomposition of the corresponding N point functions of flux creating operators.  The states with flux can be confined.  Or possibly the states with flux can be screened.  

The interaction between externally inserted charges can be studied by looking at the Wilson loop operator.  The corresponding interaction between vortices can be studied by looking at the 'tHooft loop operator\cite{tHooft}.  In Ref. 7,  instanton contributions to the expectation value of the Wilson loop were computed in the symmetry broken Higgs phase in the model without the Chern-Simons term.  Here topological vortex solitons exist in the spatial plane.   They can be continued to three dimensional Euclidean space as vortex loops which  are unstable instantons.  Their contribution to the path integral requires summing over all configurations of vortex loops.  Again the important contribution is essentially topological, the Wilson loop integral measures the topological linking number of the vortex loop with the Wilson loop.  

On a lattice of fixed spacing one finds the contribution of loops of length $L$ to be proportional to  $e^{-mL}\mu^L$, where $m$ is the vortex mass per unit length, and $\mu$ is an intrinsic parameter such that $\mu^L$ counts the degeneracy of loops of length $L$.  A phase transition occurs when $m-ln\,\mu\sim 0$ where vortex loops of all lengths become important.  This transition has been seen to occur in lattice simulations of the model, see e.g. 8.  The argument, weighing the effects of entropy versus Boltzmann suppression is robust.  It was used to identify the  Kosterlitz-Thouless phase transformation\cite{kt}.  In  our case it implies the existence of a change of phase from the usual Higgs phase that corresponds to a condensation of the vortex anti-vortex pairs at finite, non-zero vortex mass.  For non-interacting vortices, which is the case for the Abelian Higgs vortices, the argument is qualitatively certain.  Adding a perturbatively small Chern-Simons term cannot affect these conclusions.  The limit we are contemplating is one in which the Chern-Simons mass yields a negligible perturbation to the masses obtained by the Higg's mechanism for the gauge field and the scalar field.  The vortex loop configurations are known to be only perturbatively affected by the Chern-Simons term\cite{pk}.    As the classical configurations are perturbatively  modified, the semi-classical approximation is also expected to be robust, free of non-analytic behaviour as the Chern-Simons mass is taken to zero.  The zero order contribution to the expectation value of the Wilson loop operator will also only be perturbatively modified with the addition of the Chern-Simons term.  The conclusions made in Ref. 7 that vortex anti-vortex pairs condense at a small but finite vortex mass, will continue to be valid as a perturbative Chern-Simons term is switched on.  

A partition function of a gas of loops can be converted to an effective field theory of a scalar field called the vortex field.  Condensation of the vortex field corresponds to the ``spaghetti vacuum'' and is called topological symmetry breakdown\cite{Sam} which is identified as a different phase with unbroken symmetry of the model, distinct from the usual Coulomb phase of unbroken symmetry.  In the region where the vortex field condenses the vortex loops induce a nonperturbative logarithmic confining force between external charged particles \cite{Sam}.  If the external particles have charge $q$, the strength of the logarithmic potential is periodic in $q$ with period $e$, the charge of the elementary scalar particles in the theory, since they screen the external particles when $q$ is a multiple of $e$.  

To examine the behaviour of vortices themselves, one studies the 'tHooft loop operator\cite{tHooft}. It is defined by the action of a singular gauge transform on the fields, and at each time slice it creates a vortex anti-vortex pair.  The behaviour of the 'tHooft loop expectation value describes the inter-vortex potential.  In the absence of the Chern-Simons term, the usual perturbative contributions and vortex loop configurations give a perimeter law dependence to the 'tHooft loop, for both phases of the theory.  If the Chern-Simons term, with coefficient $\kappa$ is added what is its effect on the 'tHooft loop expectation values?  We find that the perimeter law changes yielding a long range interaction. The effects of the Chern-Simons term are the leading non-trivial contributions.  This is however perfectly consistent with the perturbative fashion in which we treat the Chern-Simons term.  The long range interaction is perturbative in the coefficient of the Chern-Simons term.  
The first problem that we face is that the Chern-Simons term is imaginary in Euclidean space while the rest of the action is real.  Since the action is complex, the Euler-Lagrange equations in terms of real fields over determine the system and there is in general no solution to the field equations in terms of real fields.  There do exist complex solutions.  These are however, perturbative in $\kappa$.  We then find that the real vortex loop configurations from $\kappa=0$  continue to be perturbatively the most relevant configurations. Indeed, the action is only modified at second order by the complex solutions because the Lagrangian density is quadratic in the gauge field that is perturbatively linearly modified by the Chern-Simons term.   The zero order part of the action simply comes from the zero order real vortex solution evaluated in the real part of the Lagrangian density, and the first order part comes from the evaluation of the Chern-Simons term, also with the zero order real vortex solution since the Chern-Simons term contains an explicit factor of $\kappa$.  It is this linear imaginary term in $\kappa$, that is responsible for the modification in the behaviour of the expectation value of the 'tHooft loop operator.  There is no non-analyticity as $\kappa$ is taken to zero, the long range potential that we find is perturbative in $\kappa$ and smoothly vanishes as $\kappa$ goes to zero.  Our approach is the same as \cite{AHPS}, who considered the effect of the Chern-Simons term to the monopole contribution in the Georgi-Glashow model.

We thus find that for the 2+1 dimensional Abelian-Higgs model the addition of the Chern-Simons term does not significantly alter the vortex loop contribution to the Wilson loop expectation value for small $\kappa$.  The behaviour of the 'tHooft loop is however significantly altered. We find that when the vortex field condenses, vortex loops give a logarithmic potential between both the external electric charges and vortices.
This result is surprising given that the photon is now massive due to the Higg's mechanism and perturbatively no long range forces exist.  

We give few technical details here due to lack of space, we refer the reader to our original publication\cite{ip}.  
We define our model with the following Lagrangian density,
\be 
{\cal L}=\left\{-\frac{1}{4}F_{\mu\nu}F^{\mu\nu} +\frac{1}{2}
D_{\mu}\phi\,(D^{\mu}\phi)^*-
\frac{\lambda}{4}(|\phi|^2-\eta^2)^2+\frac{\kappa}{4}\epsilon
^{\mu\nu\rho}A_{\mu}F_{\nu\rho} 
\right\}\;,\label{Ldens}
\ee
here $\phi$ is a complex scalar field, 
$A_{\mu}$ is a $U(1)$ gauge potential,
$D_{\mu}=\partial_{\mu}-ieA_{\mu}$,
$F_{\mu\,\nu}=\partial_{\mu}A_{\nu}-\partial_{\nu}A_{\mu}$ 
($\mu$, $\nu=0,\,1,\,2$)
with $\lambda$, $e$, $\kappa$ and $\eta^2$ coupling constants.
$\kappa$, $\lambda$, $\eta^2$ and $e^2$
each have the dimension of mass. 
When $\kappa=0$, if $\eta^2\leq 0$ the gauge symmetry is unbroken and the perturbative spectrum consists of a 
massless photon and a massive charged 
scalar particle of charge $e$. If $\eta^2>0$ the gauge symmetry is spontaneously broken and the particle content is given by a massive vector particle and a massive neutral scalar particle.   However, additional classical time independent vortex solitons exist called
Nielson-Olesen vortices. 
If the Chern-Simons term is now included, $\kappa\neq 0$, the theory is altered in that the photon is massive even 
when $\eta^2\leq 0$. The CS term automatically gives the photon a mass proportional to $\kappa$. The vortex solitons are also present. 
Gauss law implies that any vortex must have electric charge $Q$ (and vice versa): 
\be
\int d^2x\,J^0 =Q=\kappa \Phi\,,
\ee
where $\Phi=\int d^2xF_{12}$ is the flux and $J^{\mu}$ 
is the conserved electro-magnetic current. 

The Wilson loop operator is defined by
\be
W(C)=exp(iq\oint_CA_idx^i)\, ,
\ee
where $C$ is a closed oriented loop in Euclidean three-space,
$x^i(s)$, and $q$ a real number. 
The 't Hooft loop operator \cite{tHooft}, $B(C)$, is defined by its action 
on a field eigenstate $|A_i(x),\,\phi(x)>$, which transforms this state by 
a singular gauge transformation into 
$|A_i^{\Omega_C}(x),\,\phi^{\Omega_C}(x)>$.
The gauge transform  $\Omega_C(x)$ with
\be\label{omega}
\Omega_C(x)=exp(i\omega_C(x))\,,
\ee
it is defined by the following.  
If another closed curve $C'$ links with $C$ $n$ times in a certain 
direction and $C'$ corresponds to $x^i(\theta)$, $0\leq\theta\leq 2\pi$, 
then 
$\omega_C(x(2\pi))=\omega_C(x(0))+2\pi n$.
To calculate the expectation value of $B(C)$ it is necessary sum over all field 
configurations which include this infinitely thin 
flux tube, and each configuration of action $A$
is weighted with $e^{-A}$. 

The calculation of the expectation value of the Wilson loop and the 'tHooft loop proceeds exactly as in Ref. 7.  The sum over a gas of loops converts to an effective field theory, which is evaluated semi-classicially.  The important term in the effective field theory comes from the linking of the vortex loops with the Wilson loops.   The result for the expectation value of the Wilson loop is
\be\label{313}
<W(C)>=<exp[iq\oint_C A\cdot dx]>\approx exp[-(\frac{2\pi q}
{e})^2\chi_0^2\;ln(\frac{r}{r_0})]\,, 
\ee
where again $C$ is rectangular with lengths $r$ and $t$, $t\gg r$, 
with $r$ representing the separation of the electric charges, and 
$r_0$ is a cut off of the order of the vortex width.
The evaluation of the 'tHooft loop gives essentially the same result.  Now the complex part of the Euclidean action plays a crucial role. 
The Chern-Simons term implies that for each 
vortex loop $C'$ that links $n(C,C')$ times with $C$ we get a term, 
\be
exp\,\{i\kappa(\frac{2\pi}{e})^2n(C,C')\}\,,
\ee
where $2\pi/e$ is the flux carried by 
the 't Hooft loop $C$. This in turn means that 
the 't Hooft loop calculation is 
identical to the Wilson loop calculation with the replacement 
$q\rightarrow 2\pi\kappa/e$, which due to the Gauss law, is nothing else than 
the electric charge of a vortex that has flux $2\pi/e$. 
Hence in the region $\eta^2<\eta^2_{crit}$, vortex 
loops give a logarithmic contribution to the 't Hooft loop
expectation value and one 
expects logarithmic confinement of external 
vortices.

Electrically charged particles of charge $q\,\mbox{mod}\,e\neq 0$ are 
logarithmically confined in the topological symmetry breaking phase 
(or spaghetti vacuum) as discovered by Samuel \cite{Sam}. Our new 
result is that this remains true with the addition of the 
Chern-Simons term, at least for small coefficient $\kappa$. 
Additionally, this includes the logarithmic 
confinement of vortices 
since they also carry electric charge due to the Gauss law.
This result is surprising in view of the fact that all gauge fields
are massive and interactions short ranged and demonstrates the
importance of vortex configurations for the phase structure of the
theory.
\section*{Acknowledgments}
We thank NSERC of Canada and FCAR of Qu\'{e}bec for financial support. We thank L. Bettencourt, R. B. Mackenzie and V. Rubakov for useful discussions.  We also thank the organizers of the XVIII Autumn School, Topology of Strongly Correlated Systems, Lisboa, Portugal, for a  superb and interesting school.

\end{document}